\documentclass{eccomas2016}

\usepackage[square,sort,comma,numbers]{natbib}
\usepackage{float}
\usepackage{hyperref}
\usepackage{amssymb}
\usepackage{amsmath}
\usepackage{subfig}
\usepackage{color}
\usepackage{graphicx}

\makeatletter
\color{black}
\let\default@color\current@color
\makeatother

\renewcommand\refname{}

\title{A design-through-analysis approach using the Finite Cell Method (ECCOMAS CONGRESS 2016)}

\author{Benjamin Wassermann$^1$, Tino Bog$^1$, Stefan Kollmannsberger$^1$, and Ernst Rank$^{1,2}$}

\heading{Benjamin Wassermann, Tino Bog, Stefan Kollmannsberger and Ernst Rank}

\address{$^1$ Technical University of Munich\\
Arcisstr. 21, 80333 Munich, Germany\\
  e-mail: \{benjamin.wassermann, tino.bog, kollmannsberger\}@tum.de \and $^2$
 Technical University of Munich \\
 Institute for Advanced Study \\
 Lichtenbergstraße 2a, 85748 Garching, Germany\\
  e-mail: ernst.rank@tum.de }

\keywords{Constructive Solid Geometry, CSG, Finite Cell Method, FCM, Point-in-membership test}

\abstract{Modern 3D Computer-Aided-Design (CAD) systems use mainly two types of geometric models. Classically, objects are defined by a Boundary
Representation (B-Rep), where only the objects’ surfaces with their corresponding edges and nodes are stored. One disadvantage concerning a numerical
simulation is that B-Rep models are not necessarily water-tight.  These 'dirty geometries' cause major difficulties in computational analysis because even basic geometric operations such as point-in-membership tests fail, not to mention meshing as required by classical boundary conforming finite element
methods.  Alternatively, objects may be represented by Constructive Solid Geometry (CSG), which is strongly related to Procedural Modeling (PM). In this
context, the model is created using Boolean operations on primitives. The modeling process is then either stored as a sequence (PM), or as a construction tree (CSG). In
contrast to B-Rep models, CSG models are intrinsically water-tight. To run a finite element simulation on a water-tight CSG model, two alternatives are
possible: (i) it can either be converted to a B-Rep-model to obtain a finite element mesh or (ii) its implicit description can be used directly by applying an embedded domain approach, like the Finite Cell Method (FCM). \\
In this contribution, we present a design-through analysis methodology using CSG and FCM. A crucial point in FCM is a fast and reliable point-in-membership test which can be directly derived from the CSG model. We present the outline of the modeling approach, the realization of the point-in-membership test as a
sequence of CSG-operations, and discuss advantages and limitations on complex models of relevance in mechanical engineering.}

\begin{document}

\section{INTRODUCTION} \label{sec:intro}

Computer aided engineering often requires an an iterative process to find an optimal design. This iterative process consists of a modeling phase followed by a numerical simulation and an analysis phase. For the second phase, a common choice is the classical finite element method (FEM) in which the finite elements are conforming with the physical boundaries of the model. An estimation of the relative time required in a representative design process at Sandia National Laboratories~\citep{Cottrell2009} has shown, that the transition from the geometric model to the simulation model causes more than 80 \% of the engineering effort. \\
Various methodologies have been developed to overcome the difficulties involved in this transition process. The most prominent method is Isogeometric Analysis (IGA) as proposed by Hughes et al. \citep{Hughes2005}. IGA aims at bridging the gap between the CAD model and computational analysis by using the same shape functions in CAD and FEA. To this end, B-Splines and NURBS are used. These functions offer several desirable properties such as the possibility of straightforward refinements in grid size and polynomial degree and the possibility to control the continuity within a patch. Furthermore, as B-Splines and NURBS are functions of higher order, they offer the potential to deliver high convergence rates in case of smooth solutions of the underlying problem.\\ 
Geometric models in CAD systems are often described using a boundary representation (B-Rep)~\citep{Foley1997}. Here, IGA was first applied to surface bodies, which where made up of several conforming two-dimensional B-Spline or NURBS patches. More complicated topologies are usually generated by trimming, which may lead to non-water tight geometric models. Remedies for this problem range from classic re-parametrization \citep{Piegl1997} to the use of T-Splines~\citep{Bazilevs2010a}. B-Rep solids still pose challenges for IGA, since they are defined by a collection of their bounding surfaces. Hence, the B-Rep does not provide three-dimensional patch to directly discretize the volume.\\
However, B-Rep is not the only possible way to represent geometries. Constructive Solid Geometry (CSG) \citep{Requicha1977} expresses the underlying construction process by combining simple solid primitives with Boolean operations. Many modern CAD systems use a hybrid representation combining B-Rep and constructive solid geometry (CSG)~\citep{Gomes1991}. In this context the B-Rep model provides the additional information necessary e.g. for visualization purposes. From a design point of view, CSG offers a more intuitive approach of geometric modeling. Additionally, CSG can efficiently be used for parametric and feature based design~\citep{Shah1995} for which a description of the construction history, dependencies and constraints is mandatory.
At first sight, IGA seems to be more closely related to B-Rep models. However, even before the IGA idea became popular, Natekar et al.~\cite{Natekar2004} proposed a method to combine spline-based element formulations with two-dimensional CSG model descriptions. In contrast to the approach presented in the contribution at hand, heavy use is made of an explicit representation of boundaries and a decomposition into sub-domains. Recently, Zuo et al. \citep{Zuo2015} proposed an approach in which each CSG primitive is treated separately. The resulting sub-domains are then coupled with the Mortar method. This poses the difficulty that an explicit boundary representation needs to be set up also for inter-subdomain boundaries to span the Mortar boundary.\\
In parametric modeling a change of parameters, or constraints has hardly any impact on the CSG model, but may require a complete reconstruction of the entire corresponding B-Rep model. Hence, a simulation technique is desirable, which uses the explicit description of volumes by CSG as much and its B-Rep representation as little as possible. To this end we propose a combination of CSG and the Finite Cell Method. We denote our approach as 'design-through-analysis' as it allows, like IGA, a very close interaction of the (geometric) design process and the (numerical) analysis, where an engineer can immediately investigate consequences of a variation of the geometric design on the mechanical behavior of a structural object. \\
The Finite Cell Method (FCM) was first proposed by Parzivan et al.~\citep{Parvizian2011}. The FCM is a high-order fictitious domain method, which embeds an arbitrary complex geometry into an extended domain which can easily be meshed by a Cartesian grid. The complexity of the geometry is handled only on the integration level. This  renders the method very flexible as
the only information FCM needs from the CAD model is a reliable and robust point-in-membership test, i.e. whether an integration point lies inside or outside of the physical model. This point-in-membership test is directly provided by the CSG model description. The interplay between CSG and FCM was already investigated for simple primitives and proved to be a "very accurate and efficient method for analyzing trimmed NURBS patch structures" \citep{Rank2012}. The goal of the present paper is to extend the combination of FCM and CSG to more complex geometric models and to solid construction processes of industrial relevance. \\
This paper is organized as follows: In Section 2 a short overview on geometric representations and the Finite Cell Method is given. In Section 3 the relevant methods for the combination of CSG and FCM are presented. Section 4 provides an example showing the relevance and potential for practical applications.

\section{OVERVIEW} \label{sec:overview}
This section describes concepts of geometric modeling and the Finite Cell Method in more detail.

\subsection{Geometric Modeling}
\begin{figure}[H]\centering
	\includegraphics[width = 12 cm]{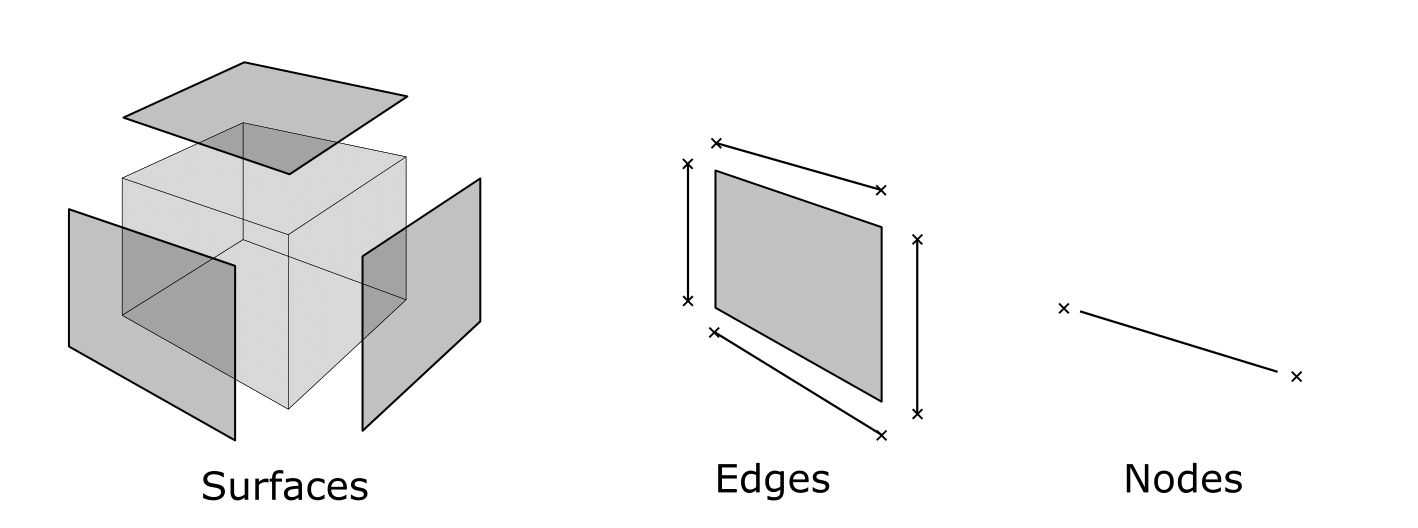}
	\caption{Boundary Representation}
	\label{fig:BRep}	
\end{figure}

Modern three-dimensional CAD systems use mainly two different types of geometry descriptions. One of them is the Boundary Representation (B-Rep), where only the objects' surfaces with their corresponding edges and nodes are stored (see figure \ref{fig:BRep}) \citep{Bungartz2004}. Although B-Rep has several advantages, such as the direct access to the surfaces, it has also some disadvantages especially with respect to a subsequent numerical simulation. B-Rep models are not necessarily water-tight. Therefore, for these invalid solids, even basic topological operations such as a point-in-membership test fail.

\begin{figure}[H]\centering
	\includegraphics[width = 12 cm]{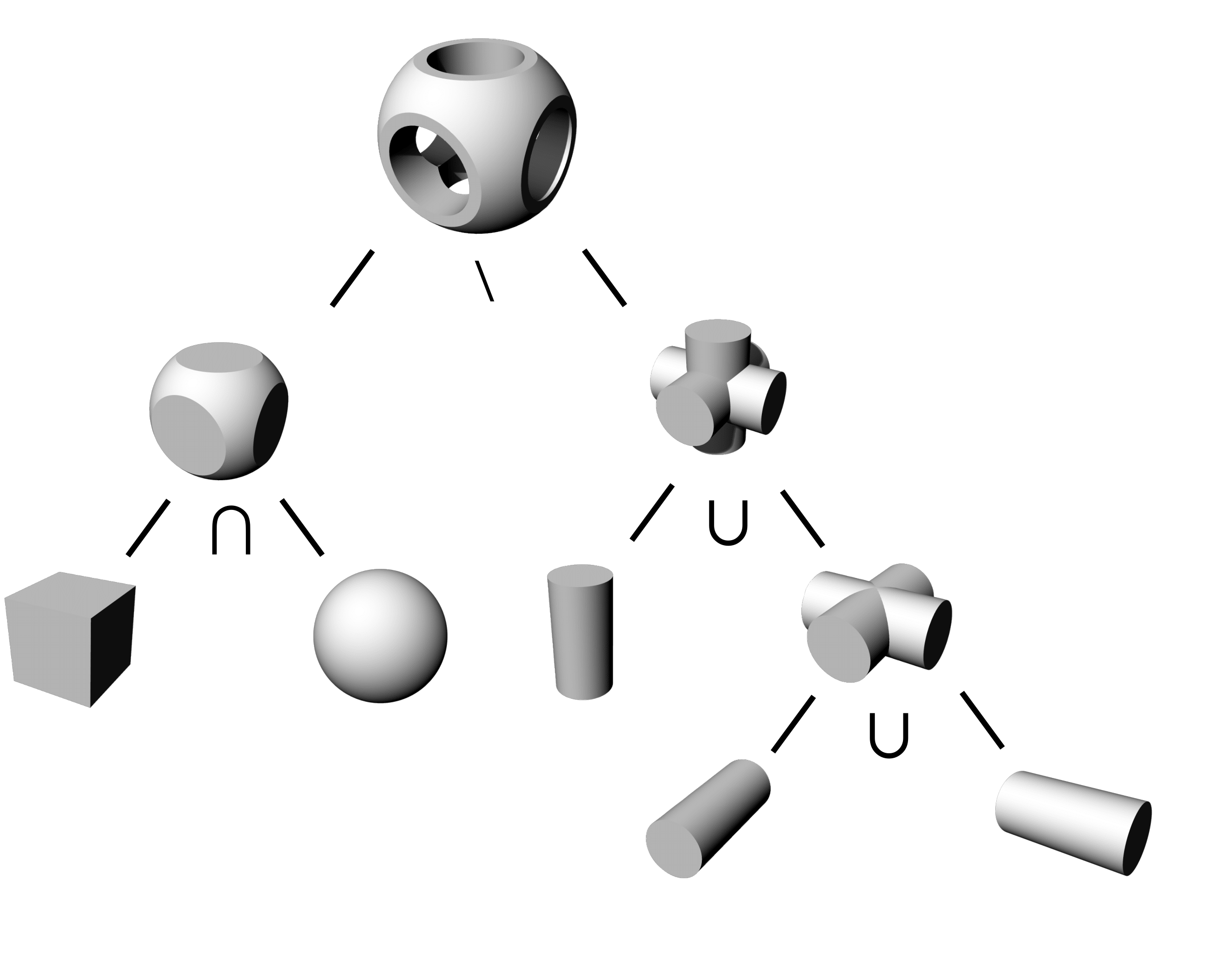}
	\caption{CSG Tree with the three Boolean operations: union $\cup$, intersection $\cap$, difference $\backslash$ on primitives}
	\label{fig:CSG}	
\end{figure}

Alternatively a 3D object is often described as a procedural model. Procedural modeling is strongly related to Constructive Solid Geometry (CSG). In CSG a 3D object is created out of a set of primitives, such as cubes, cylinders, cones, spheres, etc. These primitives are combined with the three Boolean operations: union, intersection and difference. The resulting CSG object is stored implicitly as a CSG tree (see figure \ref{fig:CSG}  ).

\begin{figure}[H]\centering
	\includegraphics[width = 7 cm]{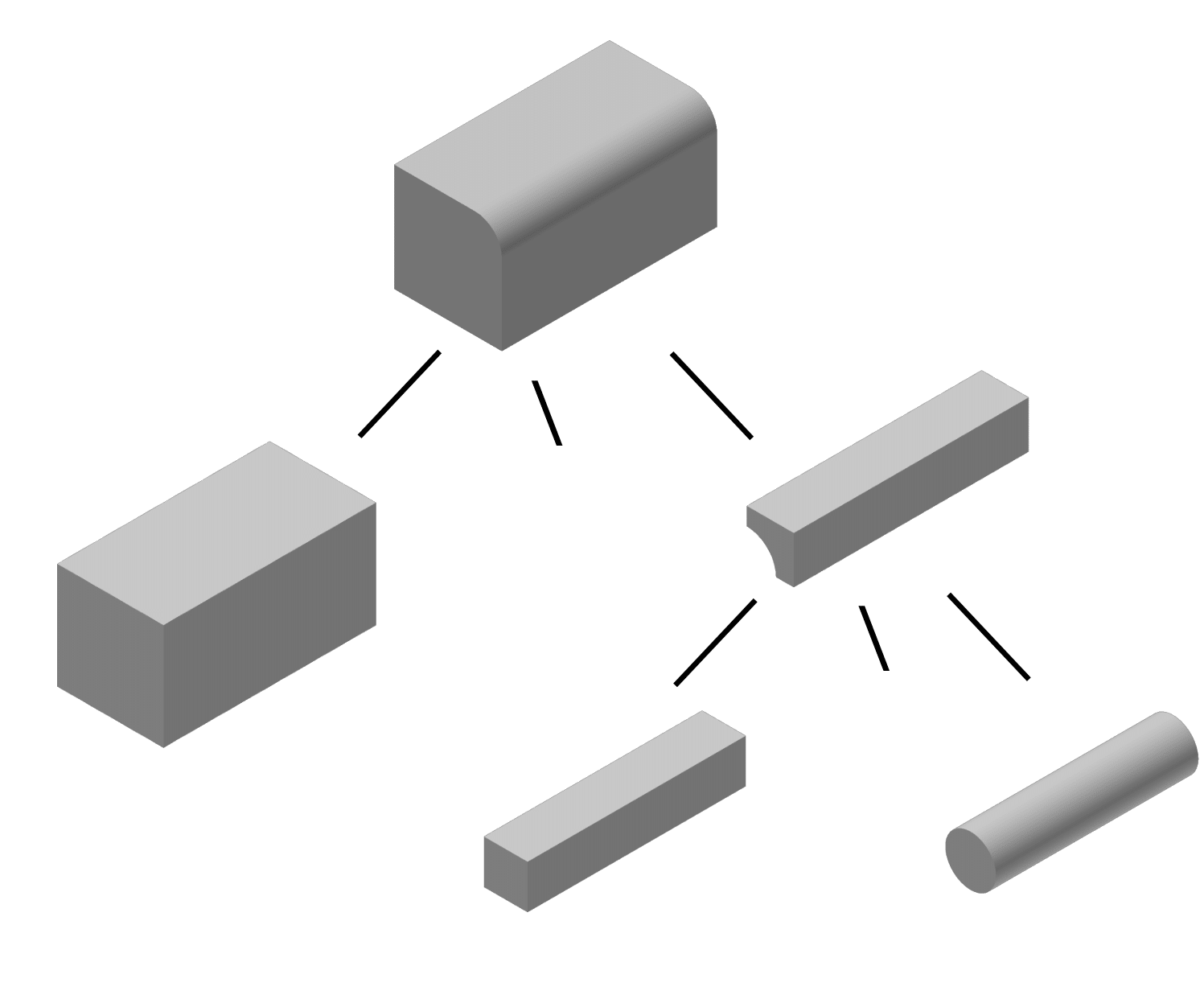}
	\caption{ Extended operations can be expressed by the classical Boolean operations: union, intersection, difference. The example shows filleting an edge.}
	\label{fig:ExtendedOperation}	
\end{figure}

\begin{figure}[H]\centering
	\includegraphics[width = 16 cm]{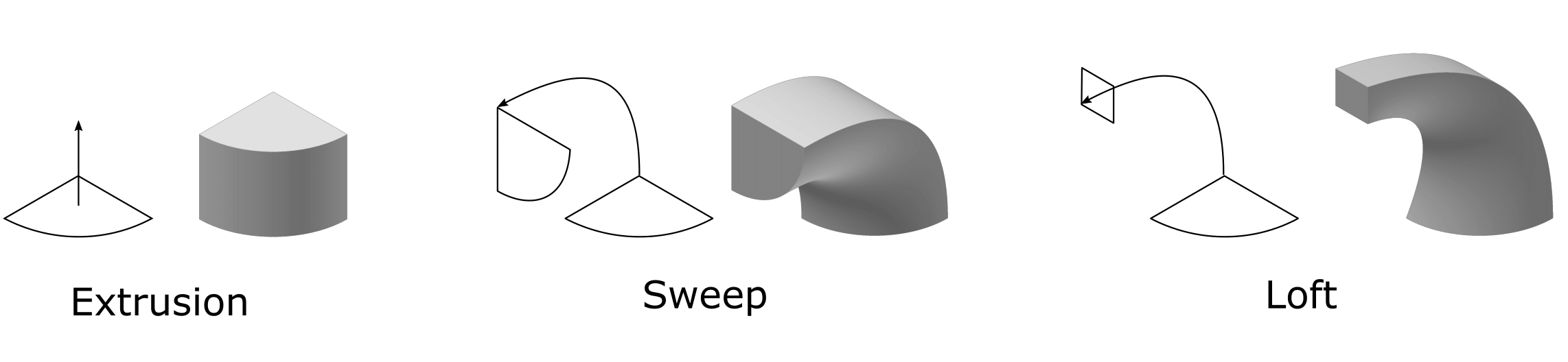}
	\caption{Extended Primitives: Extrusion, Sweep, Loft}
	\label{fig:ExtendedPrimitives}	
\end{figure}

Procedural modeling extents CSG modeling with additional operations and primitives. Extra operations, such as chamfer, fillet, drilling a hole, draft, etc. are in fact just a sequence of the original three Boolean operations, which are summarized for convenience (see figure~\ref{fig:ExtendedOperation}). A further extension allows the use of additional primitives such as sweeps, lofts or revolved objects (see figure \ref{fig:ExtendedPrimitives}). 

One big advantage of the procedural CSG model over the B-Rep model regarding a numerical simulation is the inherent water-tightness of the implicit model. 

\subsection{Finite Cell Method}
With a valid geometric description, as provided by the CSG-procedural model, it is possible to obtain a mesh on which a numerical simulation can be performed. However, mesh generators rely on a explicit (watertight) surface description of the entire boundary. This conversion may be error prone and must be conducted after each change of model. Alternatively, we aim at directly using the implicit description of the volume during the simulation process. \\
The Finite Cell Method is perfectly suited for this purpose, since it does not need to deal with complex geometries on the mesh level. The relevant geometric information is requested on the integration level, where for each integration point a point-in-membership test is performed. Points lying outside of the physical domain are penalized with a small factor \citep{Duster2008}. 

\subsubsection{Weak form}

Consider a linear-elastic problem on a physical domain $\Omega_{phy} $ with the boundary $\mathrm{d}\Omega$ divided into Dirichlet and Neumann parts $\Gamma_D$ and $\Gamma_N$. By applying the principle of virtual work the weak form of the underlying partial differential equation reads

\begin{equation}
	\mathcal{B}(\mathbf{u},\mathbf{v}) =  \int_{\Omega_{phy} } \nabla \mathbf{v} : \mathbb{C} : \nabla \mathbf{u}\; \mathrm{d}\Omega 
\end{equation}
for the inner work and
\begin{equation}
	\mathcal{F}(\mathbf{v}) =  \int_{\Omega_{phy} } \mathbf{b} \cdot \mathbf{v}\; \mathrm{d}\Omega  + \int_{\Gamma_N} \mathbf{\hat{t}} \cdot \mathbf{v}\; \mathrm{d}\Gamma
	\label{eq:linFunctional}
\end{equation}
for the external work, where $\mathbf{u}$ is the displacement, $\mathbf{v}$ the test function and $\mathbb{C}$ the elasticity tensor. $\mathbf{b}$ and $\mathbf{\hat{t}}$ denote the body load and the prescribed boundary
traction applied on the Neumann boundary, respectively.

\subsubsection{Concept of FCM}
\begin{figure}[H]\centering
	\includegraphics[width = 16 cm]{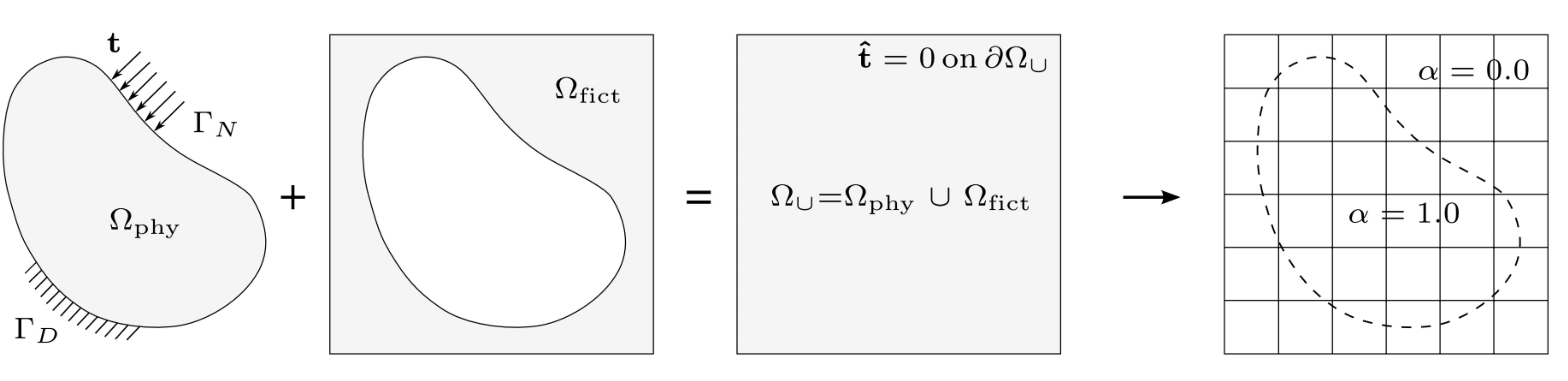}
	\caption{ Concept of Finite Cell Method, taken from \citep{Schillinger2012b}}
	\label{fig:FCM}	
\end{figure}

In FCM the physical domain $\Omega_{phy}$ is extended by a fictitious domain $\Omega_{fict}$ in such way that the resulting domain $\Omega_{\cup}$ has a simple shape and can thus be meshed easily. (see figure \ref{fig:FCM}). \\

The weak formulation is modified by defining it over the extended domain $\Omega_{\cup}$. Additionally, the virtual work terms are multiplied by a scalar field $\alpha(x)$:  \citep{Duster2008}
\begin{equation}
	\mathcal{B}_e(\mathbf{u},\mathbf{v}) =  \int_{\Omega_{\cup}} \nabla \mathbf{v} : \alpha\mathbb{C} : \nabla \mathbf{u}\; \mathrm{d}\Omega
\end{equation}
\begin{equation}
	\mathcal{F}_e(\mathbf{v}) =  \int_{\Omega_{\cup}} \alpha\mathbf{b} \cdot \mathbf{v}\; \mathrm{d}\Omega+ \int_{\Gamma_N} \mathbf{\hat{t}} \cdot \mathbf{v}\; \mathrm{d}\Gamma
\end{equation}
with $\alpha$ defined as:
\begin{equation}
\alpha = 
\left\lbrace \begin{matrix}
	1  \\
	10^{-q}
\end{matrix}
\right. \quad \quad
\begin{array}{l}
	\forall \mathbf{x} \in \Omega_{phy} \\
	\forall \mathbf{x} \in \Omega_{fict}
\end{array}
\end{equation}

To minimize the influence of the fictitious domain, while not obtaining a singular system, $q$ is typically set in the range of 5 to 10. The extended computational domain $\Omega_{\cup}$ is discretized by high-order finite elements, called cells. Current implementations use Lagrange polynomials, B-Splines or integrated Legendre polynomials \citep{Stein2004a}. Further, for an accurate integration of the bi-linear form, adaptive schemes are employed as presented e.g. in \citep{Schillinger2011}. Recently, the discretizational framework of the finite cell method has also been extended to hierarchic refinements~\citep{Schillinger2012}.

\subsubsection{Boundary conditions}
In FCM the boundaries of the physical domain $\Omega_{phy}$ typically do not coincide with the boundaries of the cells in the extended domain $\Omega_{\cup}$. This requires an enforcement of Neumann and Dirichlet boundary conditions in a weak sense. \\
Inhomogeneous Neumann conditions can be applied by integrating the prescribed traction forces $\mathbf{\hat{t}}$ along the boundary $\Gamma_N$ (see equation \eqref{eq:linFunctional}). Here, an explicit description of the boundary must be available. However, as the continuity requirements for this integration are much lower than for finite element meshes, the surface description may even be non-water-tight. \\
Dirichlet boundary conditions can be enforced weakly using methods like the penalty method~\citep{Duster2008}, Nitsche's method~\cite{Ruess2013}, Discontinuous Galerkin methods~\cite{Kollmannsberger2015} or Lagrange multipliers~\citep{Burman2010}. An explicit surface description must be available. The continuity requirements of that surface depend on the chosen method. 

\section{METHODS} \label{sec:method}

\subsection{Point-in-membership test}
The FCM performs a point-in-membership test on the integration level i.e. at each Gaussian point. The only information which is needed from the geometric model is a reliable and fast statement, if a point lies inside the physical domain $\Omega_{phy}$ or inside the fictitious domain $\Omega_{fict}$. Classically this test is done on a B-Rep model casting a ray from the concerned point into an arbitrary direction. All intersections with the boundary are then counted. The point lies inside the domain, if the number of intersections is odd, and outside otherwise. This classic test has the drawback that it fails for non-water tight B-Rep models and its computational effort rises with the complexity of the surface description, i.e. a model consisting of many non-planar surfaces or highly resolved surface triangularizations.\\
By contrast, a point-in-membership test can be performed much faster on a CSG tree. Here, the root element is queried, which forwards the request to (a selection of) its children. Since the tree is build from primitives the individual tests are very cheap. Therefore, the complexity is at its worst proportional to the number of bodies involved which is in general orders of magnitude lower than the number of surfaces for all practical applications. 

\subsection{Point-in-membership test on primitives}
For classical primitives a simple analytical function is available. Consider a primitive $\mathbf{B}_i$ which is created axis-aligned on the $x-y$ plane and assume that we define each primitive as a closed body, i.e. the boundary is included in the body. The test whether a point $\mathbf{P} = \{x,y,z\}$ is inside a primitive reads as follows for a:

\begin{itemize}
	\item \textbf{Sphere} with center point $\mathbf{C}_{Shpere}$ and radius $r_0$
	\begin{equation}
		\mathbf{P} \in \mathbf{B}_{Sphere} \quad iff \quad \quad  ||\overline{\mathbf{PC}}_{Sphere}||_2  \leq r_0,
	\end{equation}

  \item for a \textbf{Cuboid} defined by two corner points lying on its diagonal $P_{start} = [x_s,y_s,z_s]$ and $P_{end} = [x_e,y_e,z_e]$	
		\begin{equation}
		\mathbf{P} \in \mathbf{B}_{Cuboid} \quad iff \quad \quad  x \in [x_s, x_e] \wedge  y \in [y_s, y_e] \wedge z \in [z_s, z_e],
	\end{equation}
	
	\item and for a \textbf{Cylinder} defined by its center point $\mathbf{C}_{Cylinder} = \{x_c, y_c, z_c \equiv 0\}$, radius $r_0$, and height $h_0$ 
	\begin{equation} 
		\mathbf{P} \in \mathbf{B}_{Cylinder} \quad iff \quad \quad   ||\overline{\mathbf{\tilde{P}C}}_{Cylinder}||_2  \leq r_0 \wedge z \in [0, h_0]
	\end{equation}	
where point $\mathbf{\tilde{P}} = \{x,y,0\}$ is the projection of point $\mathbf{P}$ onto the $x-y$ plane.
\end{itemize}

There are also fast analytical solutions for other primitives like cones, pyramids, tori and frustums. However, it is not likely that these primitives are only constructed axis-aligned on the $x-y$ plane. Therefore, at a suitable position a local orthonormal coordinate system $A$ is constructed. It is spanning a work plane on which the respective primitive can be constructed. To perform a point-in-membership test, the point of interest $P$ needs to be mapped from the Cartesian space $E$ to the local base $A$

	\begin{equation} 
		\mathbf{P_A} = Q_{EA} \cdot \mathbf{P_E} + \mathbf{v}
	\end{equation}	
with $v$ the transposition vector between the center points of the Cartesian and local basis system
	\begin{equation} 
		\mathbf{v} = \mathbf{C_A} - \mathbf{C_E} = \mathbf{C_A}
	\end{equation}	
and
	\begin{equation} 
		 Q_{EA} = \begin{bmatrix}
		 A_{1x} & A_{2x}  & A_{3x}\\
		 A_{1y} & A_{2y}  & A_{3y}\\
		 A_{1z} & A_{2z}  & A_{3z}
		 \end{bmatrix}
	\end{equation}
with $\mathbf{A_i}$ being the base vectors of the local basis system $A$.

\subsection{Point-in-membership test on sweeps}
\begin{figure}[H]\centering
	\includegraphics[width=16 cm]{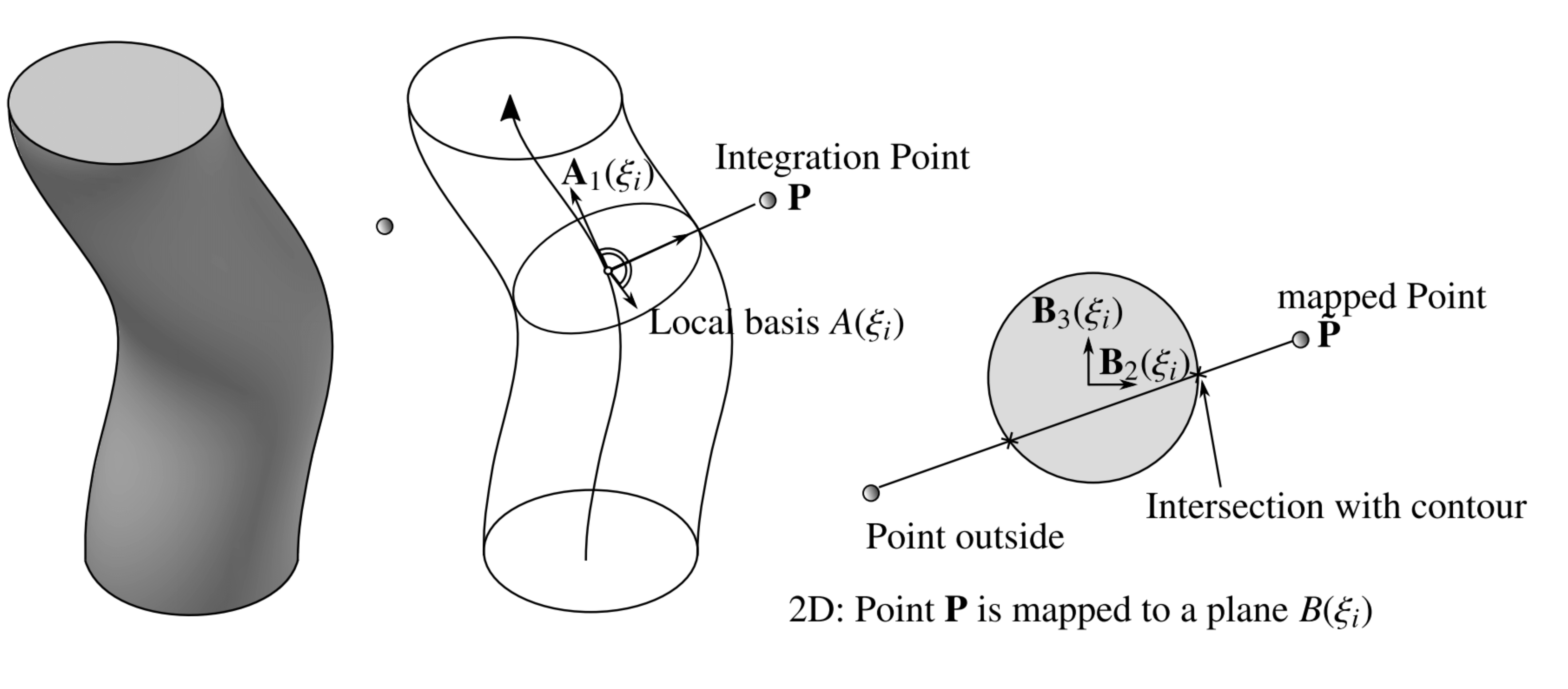}
	\caption{Point-in-membership test on intermediate sketch of a loft}
	\label{fig:Sweep}	
\end{figure}

Point-in-membership tests are more involved for "primitives" generated by sweeps or lofts. Typically, no analytic solution is available for those primitives. Nevertheless, it is possible to perform a fast, reliable test also on these bodies. The basic idea is to reduce the dimension of the problem. For this consider the set-up of a sweep. A sweep consists of a 2D sketch, which is moved along a sweep path. (see figure \ref{fig:Sweep}) \\

For the simple case that (i) the sweep path is orthogonal to the sketch plane of the starting sketch and (ii) the local basis system follows the tangent of the path, a point-in-membership carries out the following steps:
\begin{itemize}
	\item The closest point $\mathbf{{C}}(\xi_{cp})$ on the sweep path is computed either analytically or, if not possible, using Newton's method:
	\begin{equation}
			f(\xi) = \mathbf{\dot{C}}(\xi) \cdot  (\mathbf{P}- \mathbf{C}				(\xi)) \equiv 0	
	\end{equation}
	
	\begin{equation}
			\xi_{i+1} = \xi_{i} - \frac{f(\xi_i)}{f'(\xi_i)} = \xi_i - 					\frac{(\mathbf{\dot{C}}(\xi_i) \cdot  (\mathbf{P}- \mathbf{C}				(\xi_i))}{\mathbf{\ddot{C}}(\xi_i) \cdot  (\mathbf{P}- 						\mathbf{C}(\xi_i)) + |\mathbf{\dot{C}}(\xi_i)|^2}
	\end{equation}
	where $\mathbf{{C}}$ denotes an arbitrary curve description with its first and second derivative ($\mathbf{\dot{C}}$, $\mathbf{\ddot{C}}$) and $\mathbf{P}$ being the point of interest.

	\item On the closest point an auxiliary plane $WP(\xi_{cp})$ is created. For this purpose, the tangent vector at $\mathbf{{C}}(\xi_{cp})$ is evaluated and a local base system is created using e.g. the Frenet base \citep{Sternberg1964}, or a base system, where one base vector is always in a plane parallel to an arbitrary plane. 
	
	\item The point of interest $\mathbf{P}$ is mapped to the local coordinate system of the plane to obtain $\mathbf{\tilde{P}}$. 
	
	\item A point-in-membership test using a ray-test is performed in 2D with the contour line. In case of a sweep a ray cast on the initial swept object, i.e. the two-dimensional curve is carried out.
\end{itemize}

The fall back to a point-in-membership test in two dimensions poses a draw-back at first sight. However, a two-dimensional ray test is much simpler and more robust to implement than a general three-dimensional one. Clearly, a necessary pre-requisite is that the curve is closed to produce a closed object during the sweep operation. Again, this proves to be much simpler than assuring closed surfaces in three dimensions. 

\section{EXAMPLE} \label{sec:numExp}

\begin{figure}[H]\centering
	\includegraphics[width=8 cm]{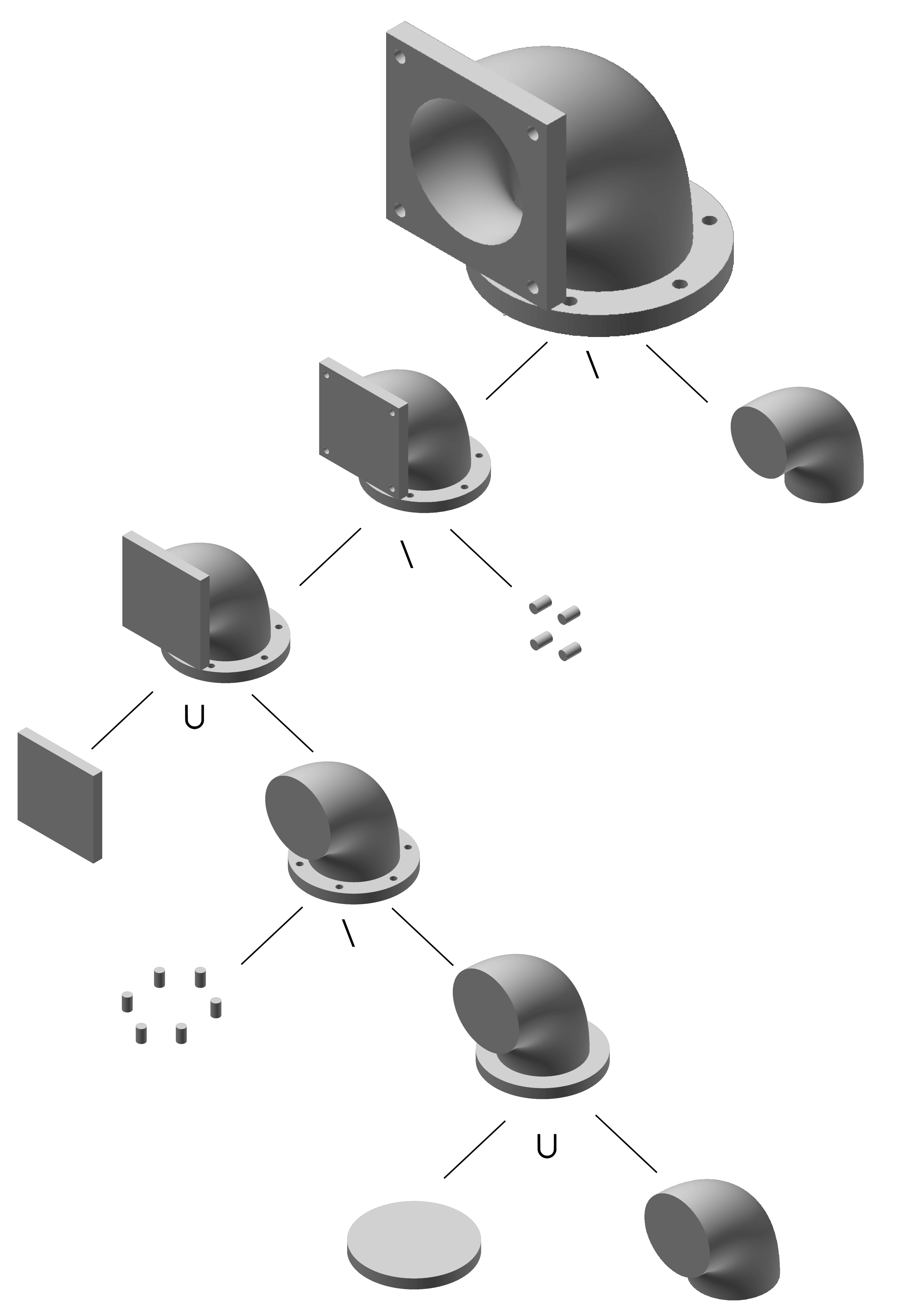}
	\caption{ CSG tree of example}
	\label{fig:SweepCSG}	
\end{figure}

The following example was constructed as a procedural model and then transformed to a CSG tree (see figure \ref{fig:SweepCSG}). It combines several simple primitives and two sweeps along a B-Spline ($p=2$) curve. The (round) base plate was fixed and a predefined deflection $\mathbf{\hat{u}} = 1$ was applied onto the left (quadratic) base plate. Homogeneous Neumann boundary conditions were applied to all other surfaces.

\begin{figure}[H]\centering
	\includegraphics[width=12 cm]{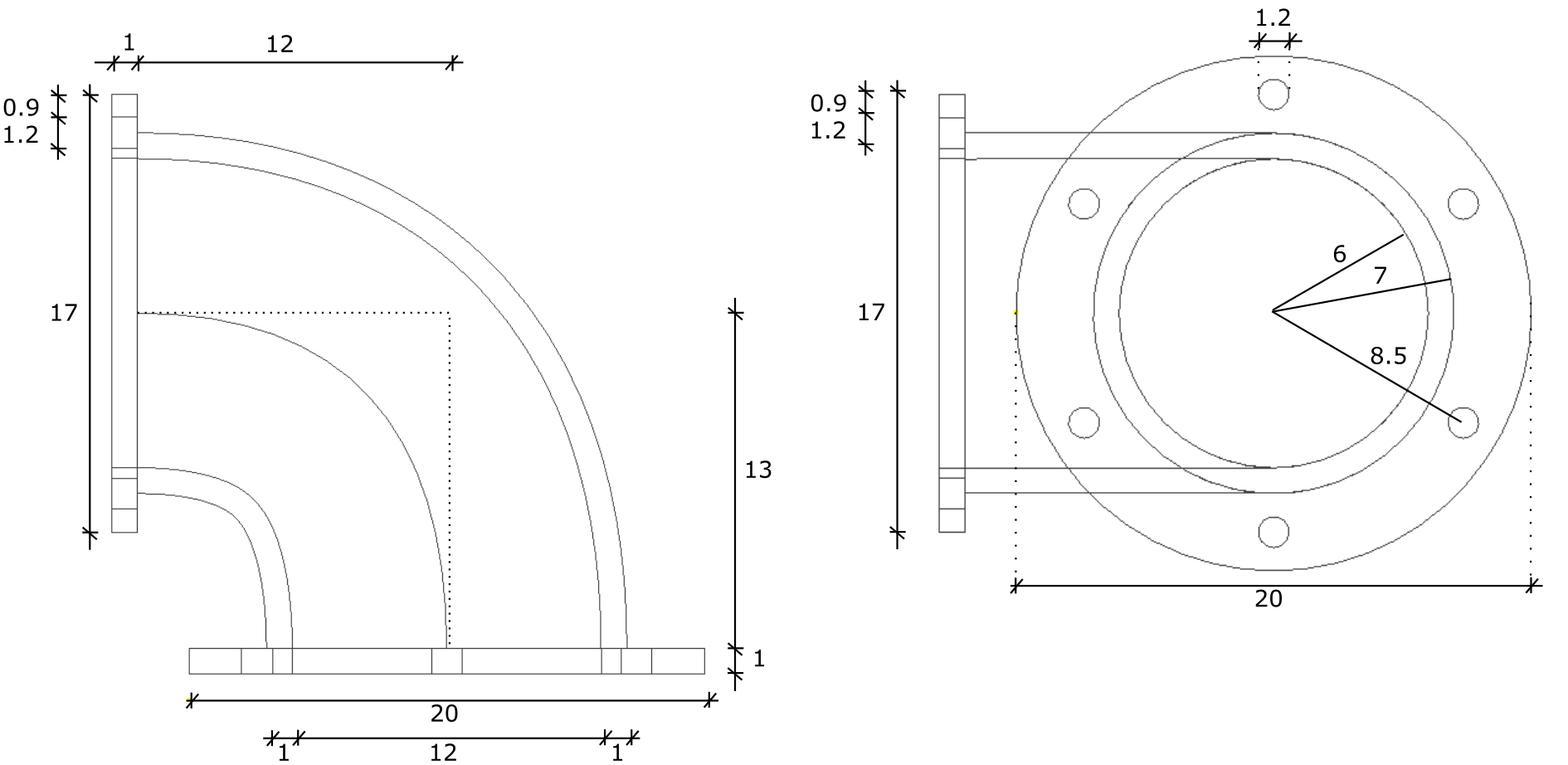}
	\caption{Dimensions of the example}
	\label{fig:SweepWire}	
\end{figure}

For the simulation in each direction 15 cells with (integrated) Legendre shape functions and a polynomial degree of $p=4$ were chosen. For a precise integration of the stiffness matrix the cells were partitioned with an octree to a maximum depth of four subdivisions. Only cells containing parts of the physical domain were considered. This reduces the number of degrees of freedom by 72 \% to 55,296 dofs.

\begin{figure}[H]\centering
	\includegraphics[width=16cm]{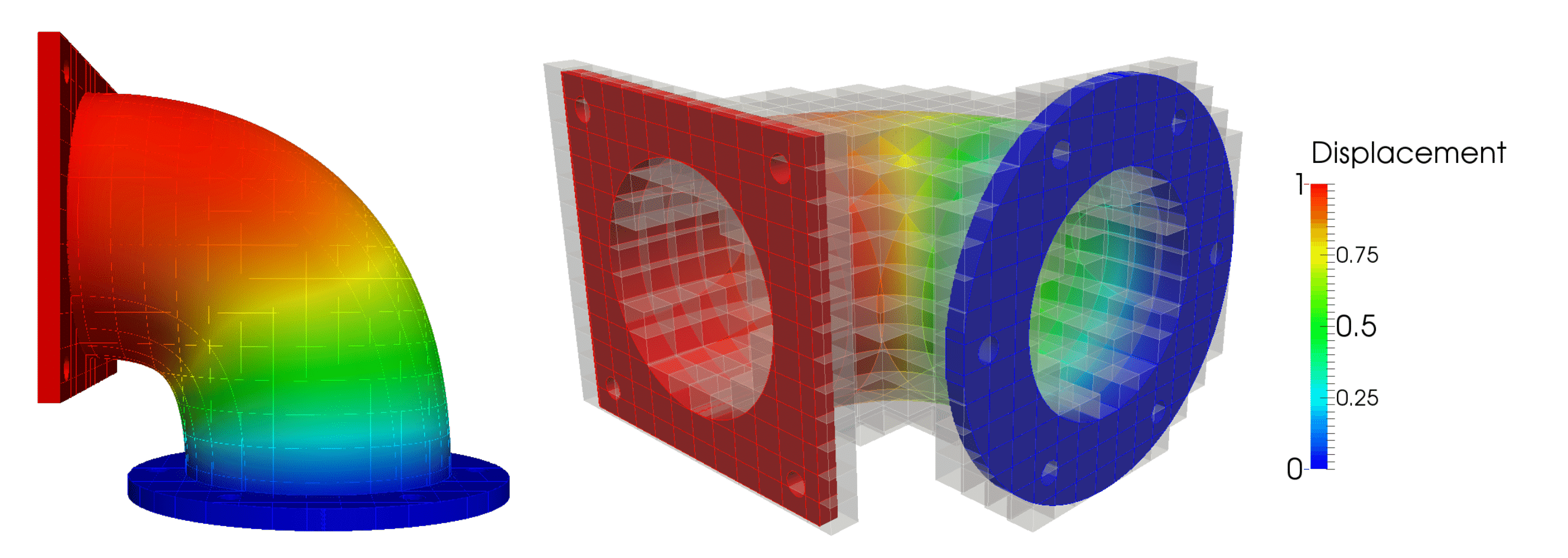}
	\caption{ Displacement }
	\label{fig:SweepDispl}	
\end{figure}

\begin{figure}[H]\centering
	\includegraphics[width=16cm]{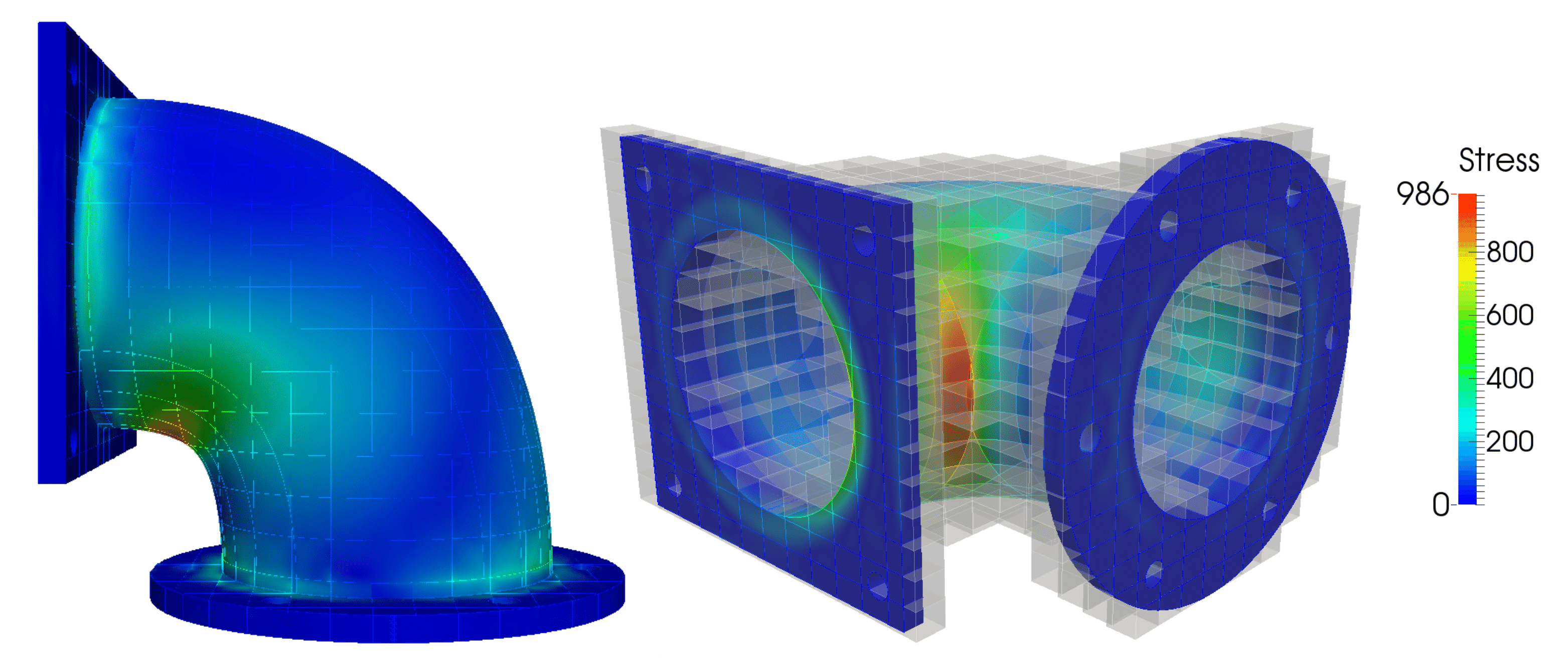}
	\caption{ Stresses }
	\label{fig:SweepStress}	
\end{figure}

Figure \ref{fig:SweepDispl} shows the finite cells embedding the structural model and the computed displacements for the example. The von Mises stresses are depicted in figure~\ref{fig:SweepStress}. These stresses provide a good overall insight into the structural load carrying behavior. Local stresses are not always fully resolved for example at re-entrant corners. A locally very accurate resolution of these singularities is possible by application of hierarchical refinements as recently developed in \citep{Zander2015}. Further extensions include the development of point-in-membership
tests for more complex CSG models such as lofts.\\
It is noteworthy that only the CSG model was used in all involved steps, i.e. from the setup of the model until the computation itself. The only point at which a conversion from the CSG-model to an explicit B-rep was carried out was for the post-processing step. Here, the marching cubes algorithm was used to derive a triangulated surface on which the results were post-processed \citep{WilliamE.Lorensen1987}. However, even this conversion is not mandatory as volumetric post-processing is a possible option as well. 

\section{CONCLUSION} \label{sec:conclusions}

An integration of the design process and numerical analysis without a complex transition like meshing is of high relevance for the industry and has been in the focus of several research groups for the last years. While Isogeometric Analysis provides an excellent method for the numerical simulation of boundary representation models and shell structures, this paper has focused on models created with Constructive Solid Geometry. A design-through analysis approach  combining CSG modelling and the Finite Cell Method (FCM) has been presented. FCM is able to use the implicit model description provided by the CSG model and hence greatly simplifies the meshing process. It was  shown that point-in-membership tests can be carried out efficiently for complex geometries like sweeps. Further steps include the development of point-in-membership tests for other primitives like lofts and sweeps with rotated sketches. 

\section*{ACKNOWLEDGMENTS}

This work has been founded by the German Research Foundation (Project RA 624/22-1). This support is gratefully acknowledged.

\vspace{0.5 cm}

\renewcommand\refname{\vskip -1cm}
\section*{REFERENCES}
\vspace{0.3cm}
\bibliographystyle{ieeetr}
\bibliography{library}


\end{document}